\documentclass[3p,times,procedia]{elsarticle}

\usepackage{ecrc}

\volume{00}

\firstpage{1}

\journalname{Physics Procedia}

\runauth{}

\jid{phpro}

\jnltitlelogo{Physics Procedia}

\CopyrightLine{2011}{Published by Elsevier BV. Selection and/or peer-review under responsibility of the organizing committee for TIPP 2011.}

\usepackage{amssymb}
\usepackage{booktabs}
\usepackage{wrapfig}

\usepackage[figuresright]{rotating}

\newcommand{\dzero}{D\O~}
\newcommand{\rdzero}{R2D\O}

\begin{document}

\begin{frontmatter}



\dochead{TIPP 2011 -  Technology and Instrumentation for Particle Physics 2011}

\title{Long-term Running Experience with the Silicon Micro-strip Tracker at the \dzero detector}


\author[primAuthor]{Andreas W. Jung$^{1,}$\let\thefootnote\relax\footnote{$^{1}$Primary author, email contact: ajung@fnal.gov.}}
\author{\\M.~Cherry, D.~Edmunds, M.~Johnson, M.~Matulik, M.~Utes, T.~Zmuda and the SMT group}

\address[primAuthor]{Fermilab, Batavia, IL, 60510, USA}

\begin{abstract}
The Silicon Micro-strip Tracker (SMT) at the \dzero experiment in the Fermilab Tevatron collider
has been operating since 2001. In 2006, an additional layer, referred to as 'Layer 0', was 
installed to improve impact parameter resolution and compensate for detector degradation 
due to radiation damage to the original innermost SMT layer. The SMT detector provides 
valuable tracking and vertexing information for the experiment. This contribution will highlight 
aspects of the long term operation of the SMT, including the impact of the silicon readout
 test-stand. Due to the full integration of the test-stand into the \dzero trigger framework, 
this test-stand provides an advantageous tool for training of new experts and studying subtle 
effects in the SMT while minimizing impact on the global data acquisition.
\end{abstract}


\end{frontmatter}


\section{Introduction}
\label{toc:intro}

\begin{figure}[ht]
    \centerline{\includegraphics[width=0.65\columnwidth]{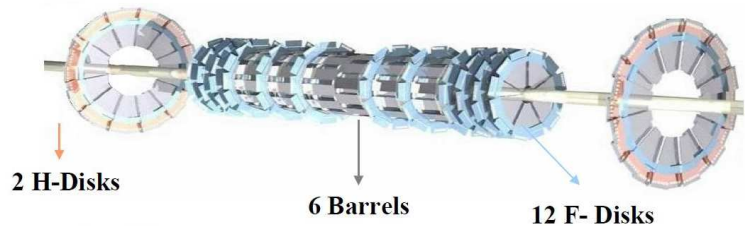}}
  \caption{\label{fig:smtlayout} The upgraded SMT detector consists of 6 Barrels, 12 'F-disks' and 2 'H-disks'. Barrels are interspersed 
with 'F-disk'. Additional 'F-disks' and 'H-disks' are placed at both ends of the detector.}
 \end{figure}

The Run II \dzero detector has been operating since 2001. It consists of two central
tracking detectors inside a $2~\mathrm{T}$ solenoidal magnet; central and forward preshower systems; liquid
argon calorimeters; and muon spectrometers including a $1.8~\mathrm{T}$ toroidal magnet. The Silicon Micro-strip Tracker (SMT) 
is part of the central tracking system of the \dzero detector and is the innermost layer of instrumentation \cite{smt}. Thus radiation damage is a potential issue and needs to be monitored and addressed \cite{zhenyu}. The SMT layout is shown in Figure \ref{fig:smtlayout}.
It consists of six barrels each with four-layers. These barrels are interspersed with six disks of small radius, so-called 'F-disks'. There are another six 'F-disks' beyond the end of the barrels. Two (originally four) large radius detectors, so-called 'H-disks', are located at the ends of the detector to enhance tracking at very large pseudo-rapidities $|\eta| < 3$. The barrels provide tracking 
for particles with high transverse momentum in the central regions $|\eta| < 1.5$, while the disk detectors allow 
for the precise reconstruction of particles traveling with pseudo-rapidity up to $|\eta| < 3$. A major SMT upgrade took place in 2006 to install an innermost layer (Layer 0) \cite{smt_l0}. This single-layer detector consists of eight barrels and was installed to mitigate the degradation of the first layer of the original SMT due to radiation 
damage. One 'H-disk' was removed from each end of the detector in 2006 to accommodate the Layer 0 readout channels. \\
There are two different types of readout chips used for the SMT: SVX-IIe \cite{svx2e} for the original SMT and SVX4 readout chips \cite{svx4} for Layer 0. The SVX-IIe readout chips are mounted on so-called HDIs (High Density Interconnects), made from Kapton flex circuits laminated to Beryllium substrates. The silicon sensors are glued to them and are referred to as 'module' \cite{smt}. There are 432 such modules for the barrel, 288 for the 'F-disks' and 96 for the 'H-disks', for a total of 5712 SVX-IIe chips installed. For L0 the HDIs are ceramic hybrids made of Beryllium oxide. Including Layer 0 there are a total of ~730k readout channels providing the largest data flow of all \dzero sub-detectors.

\section{Long-term Operational Experience}
The SMT has generally operated 24 hours per day 7 days a week since 2001, which required dedicated shift personnel and experts. In general, the operation is very stable and the response and recovery from problems is usually quick. Shift personnel are supported by on-call experts. Furthermore senior experts %
\begin{wrapfigure}{r}{0.60\textwidth}
  \begin{center}
    \includegraphics[width=0.595\textwidth]{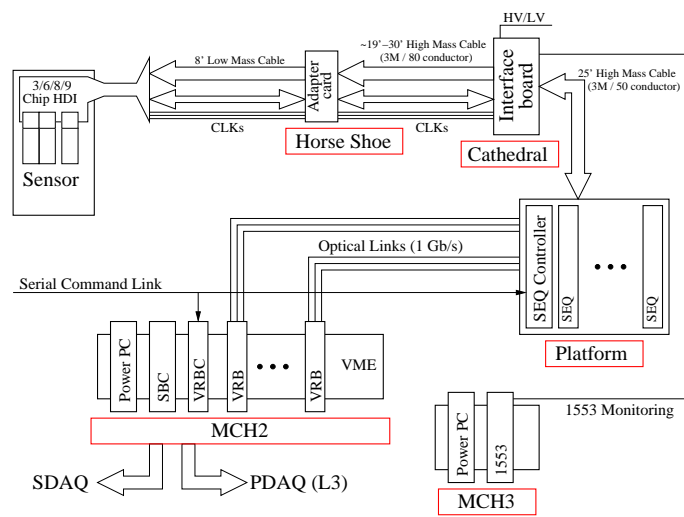}
  \end{center}
   \caption{\label{fig:hardwareChain} The hardware and readout chain of the original SMT from the sensor-HDI level to the movable counting house level (MCH) via horse shoe, cathedral and platform level.}
\end{wrapfigure}%
are available for support in various aspects. Over time many tools have been developed to monitor low and high level information such as voltages of power supplies or on-line cluster charge and size histograms as well as on-line track efficiencies. The hardware and readout chain of the SMT as sketched in Figure \ref{fig:hardwareChain} is distributed over several physical locations. These locations are not entirely accessible on a daily basis: the 'Horse shoe' and 'Cathedral' area only during longer shutdowns whereas the 'Platform' area can be accessed between stores with agreement of Tevatron operations.\\
A failure in the hardware and readout chain needs to be understood and then traced down to its physical location with the monitoring 
information at hand. In order to do that it is very important to have monitoring capabilities for low and high level information. This 
includes the monitoring of voltages and current draws of the power supplies (PS) of the detector. For example a failure of a power supply 
at the 'Platform' area typically causes lower efficiency for approximately a couple of hours until the failure is addressed. On average all power supply failures including the ones in the cathedral area compromised about 2.5\% of the collected data. The electrical crew developed a robotic remote switch '\rdzero' to switch to a spare PS within minutes for the 'Platform' PS failures. All SMT power supplies at the 'Platform' area have since been equipped with '\rdzero' units and resulting data quality losses are minimized.

\section{SMT test-stand activities}
\begin{figure}[ht]
\center
     \centerline{\includegraphics[width=1.\columnwidth,angle=0]{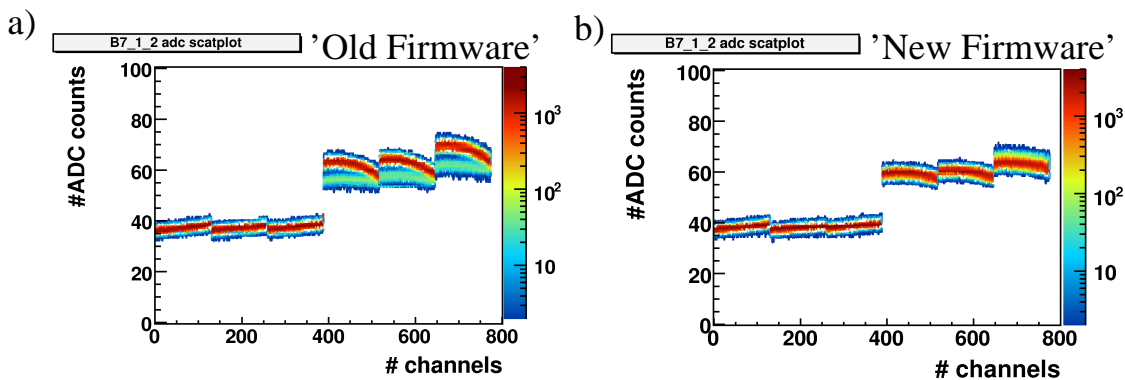}}
  \caption{\label{fig:pedestals_new} Pedestal distribution for 6 chips with 128 channels per chip. First three chips are p-side and last 
three chips are n-side. The pedestal distribution for the 'Old Firmware' is shown on the left, whereas the one for the 'New Firmware' is 
shown on the right.}
 \end{figure}

\begin{figure}[ht]
\centerline{
   \includegraphics[width=0.405\columnwidth]{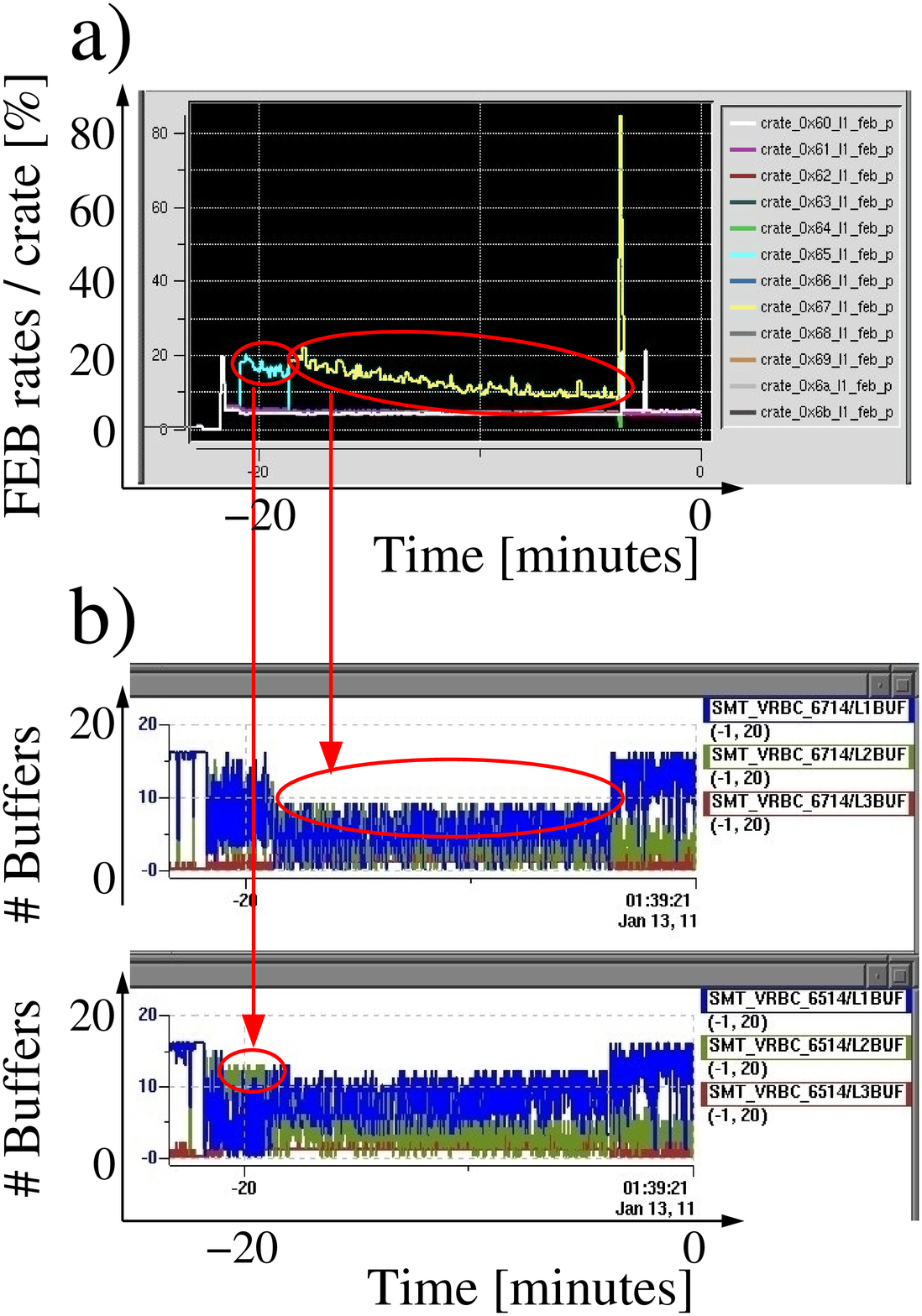}
   \includegraphics[width=0.575\columnwidth]{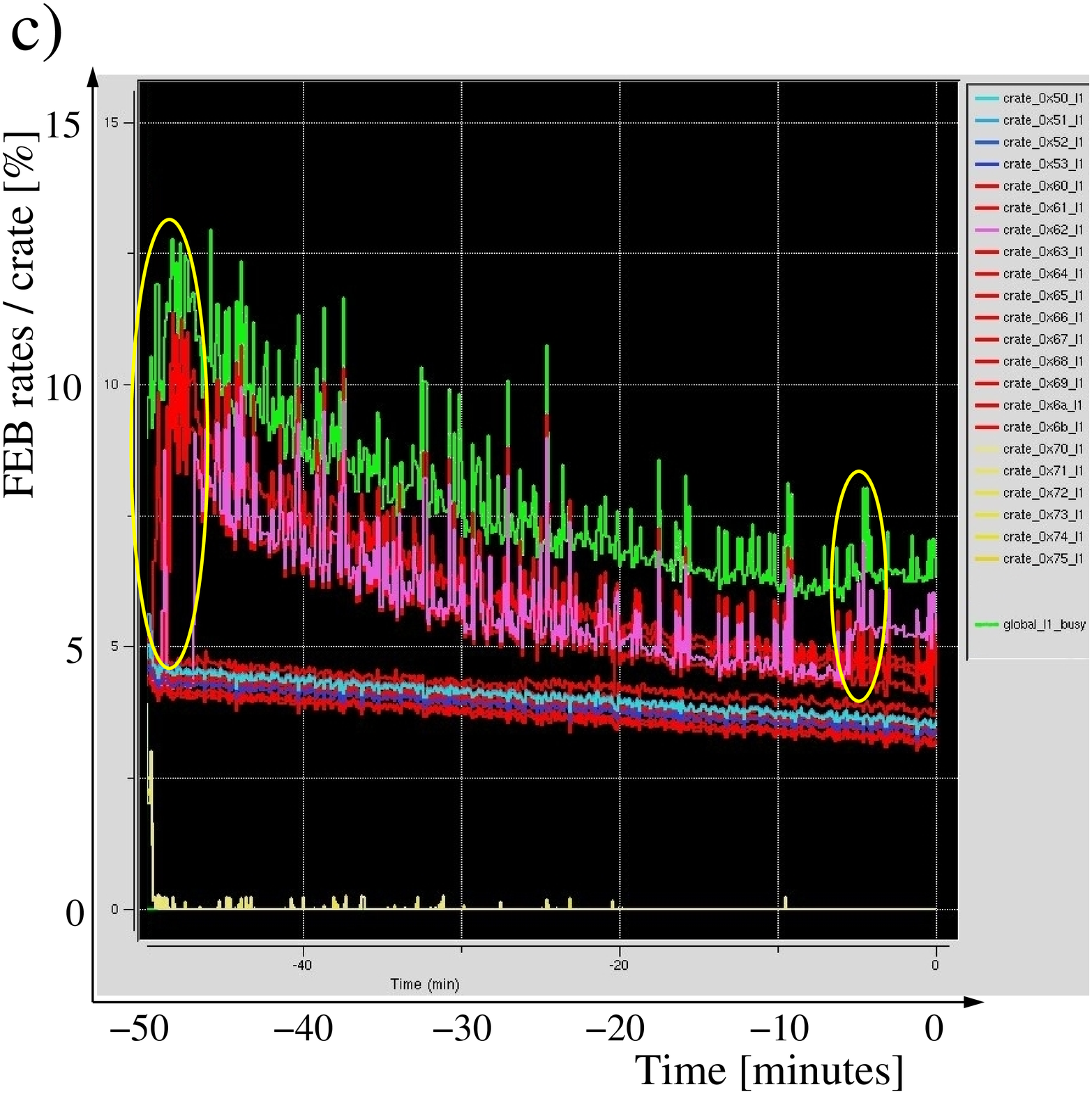}
}
  \caption{\label{fig:feb_pattern} a) shows FEB rates [\%] of all SMT crates (different colors) as a function of time without the improved buffer handling firmware. The two plots in b) show the available buffers (blue), buffers waiting for L2 (green) or L3 decisions (red) as a function of time. As an example the buffer distribution is shown for the two SMT crates which exhibit an increased FEB rate (top plot, crate 0x65 \& 0x67) as highlighted by the red ellipses and arrows. c) shows the FEB rates [\%] of various detector subsystem crates grouped by colors (SMT crates are colored in red). In addition the global \dzero L1 busy rate (green) consisting of all L1 subsystems is shown. More details in the text.}
 \end{figure}

The SMT test-stand provides a small 'copy' of the full hardware and readout chain of the SMT. In contrast to the SMT, all parts are easily accessible allowing for detailed studies of single components.
\begin{figure}[ht]
\begin{minipage}[b]{0.475\linewidth}
\centering
\includegraphics[width=0.95\columnwidth]{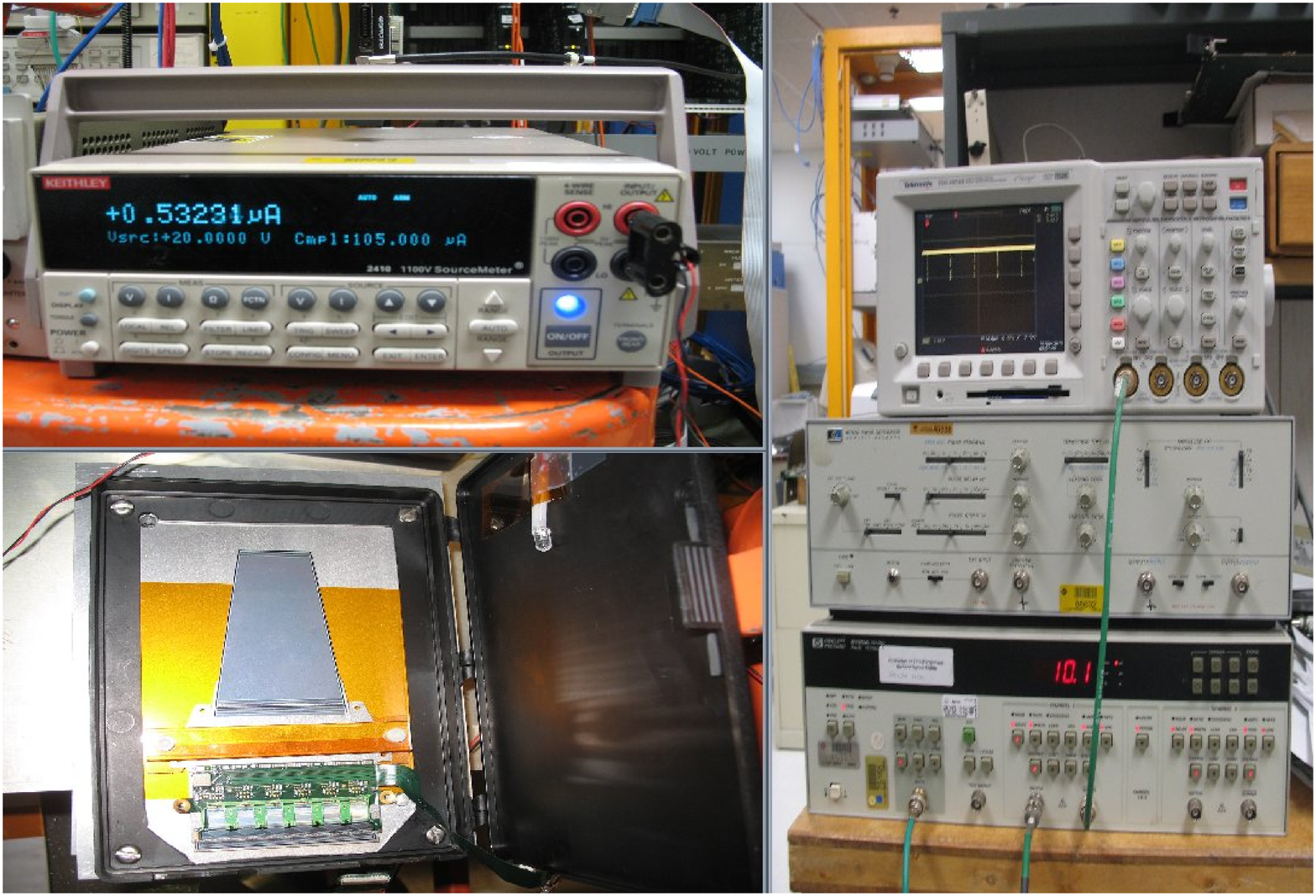}
\caption{\label{fig:teststand} The picture shows part of the SMT test-stand setup: sensor and a light emitting diode (lower left), PS (upper left), optional signal delay generator with respect to the Tevatron clock (right).}
\end{minipage}
\hspace{0.25cm}
\begin{minipage}[b]{0.475\linewidth}
\centering
\includegraphics[width=0.95\columnwidth]{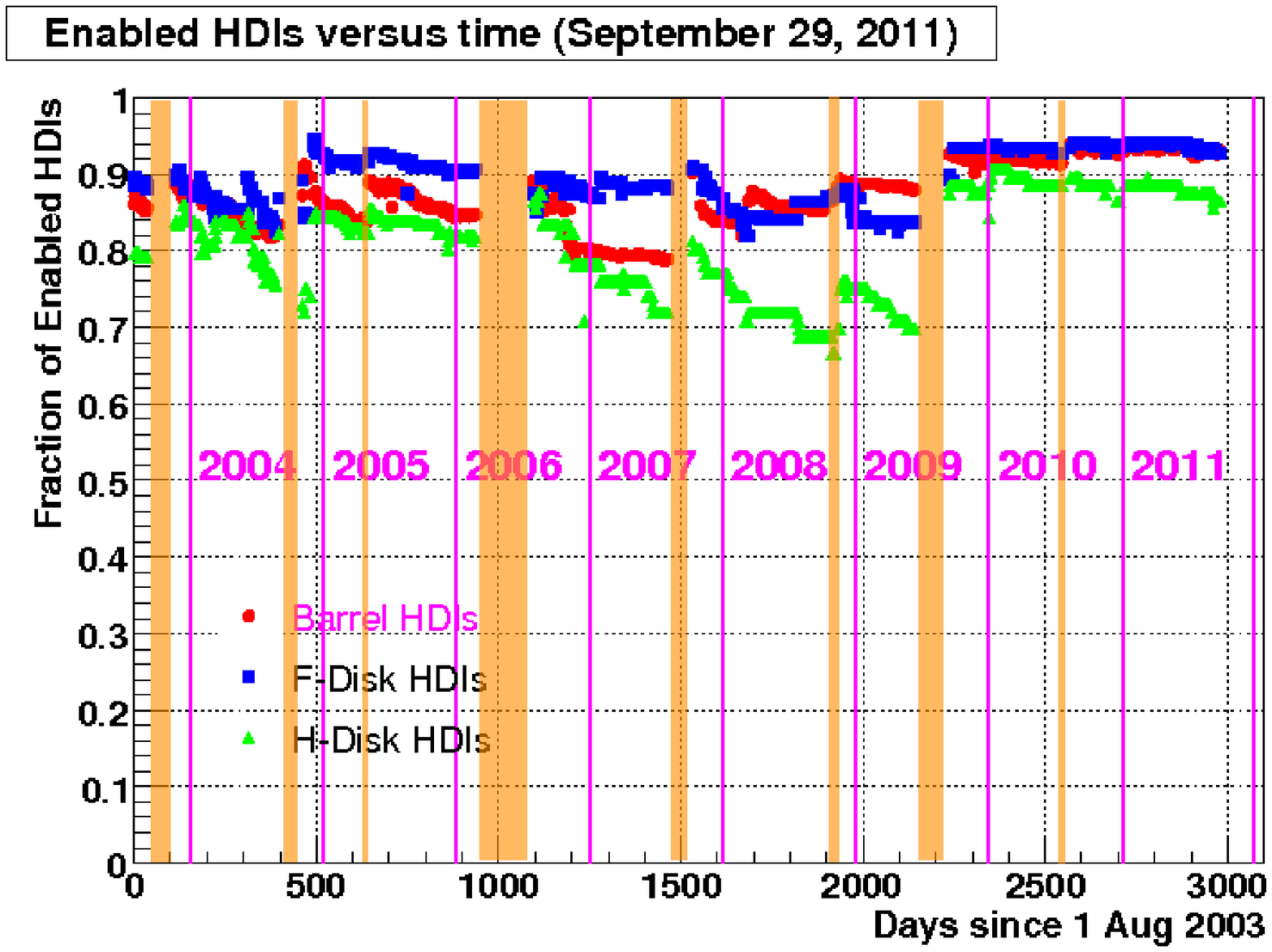}
\caption{\label{fig:enabledHDIs} The graph shows the fraction of enabled HDIs for Barrel, 'F-disk' and 'H-disk' sensors versus time. The shaded yellow bands reflect the shutdown periods of the \dzero experiment. For a more detailed explanation of the steps in the fraction of enabled HDIs see text.}
\end{minipage}
\end{figure}
Figure \ref{fig:teststand} shows parts of the test-stand setup for signal-to-noise studies with HV supply (top left), sensor
with a light emitting diode (bottom left) and optional signal delay generator with respect to the Tevatron clock (right). In general spare boards are tested at the test-stand before they are installed. This is also true during the longer shutdowns where problematic boards are replaced. During these 
shutdowns every effort is made to improve the system stability and efficiency. For a complex system like the SMT it is difficult to 
cite a single quantity characterizing global performance. A good measure is the fraction of enabled HDIs as a function of time as shown in Figure \ref{fig:enabledHDIs}. Prior to the 2009 shutdown there was a gradual decrease of this fraction due to hardware issues during continued operation. There are many different types of failures and the most common ones are individual chip failures as well as bad cable connections at various levels as given in Table \ref{tab:hdi_error} (larger steps in the fraction of enabled HDIs are explained later in the text).
\begin{table}[htdp]
\small
\begin{center}
\begin{tabular}{l l l}
\toprule
Type of HDI failure & 2009 / \# HDI & 2010 / \# HDI\\ \midrule
Adapter card & 16 & 1\\
Clock cables & 9 & 10\\
Interface Board & 15 & $-$\\
Reseating cables & $\approx$ 6 & 2\\
Bad/dead (problem inside detector) & 20 & Not re-visited\\
Disabling bad chips & 24 & 4 \\
Total \# HDI worked on & 90 & 17\\
\bottomrule
\end{tabular}
\end{center}
\caption{Detailed list of HDI defects for the years $2009-2010$.}
\label{tab:hdi_error}
\end{table}%
In order to trace such a failure to its underlying cause every failure is characterized 
and a record of previously tried interventions is maintained. The shutdown periods are highlighted with shaded bands in Figure \ref{fig:enabledHDIs}. They allowed for the time-consuming task to investigate and fix these individual failures. The efforts resulted in higher fractions of enabled HDIs after a shutdown. Another example of a sort of failure are broken wire-bonds which interrupted the distribution of the digital power lines to the readout chips. As the readout chips on an individual HDI are daisy-chained, a single chip failure caused the 'loss' of all subsequent chips of a module consisting of up-to 9 chips. The most prominent occasion occurred in late 2006 but it is likely that this sort of failure also contributed to the losses of enabled HDIs prior to that incident. The test-stand facilitated the development of a solution to this problem by using an alternative path to distribute the digital power using a special hardware board (new adapter card). Thus the initial failing chip was bypassed and the readout of the remaining chips on the module could be fully restored as implemented during the shutdown in 2007, which increased the fraction of enabled HDIs by about 10\%. Furthermore the test-stand facilitated the development of an improved sequencer firmware version as well as a modified version of the adapter card in order to fix a noise problem. Both have been installed during the shutdown in 2008 and increased the fraction of enabled HDIs. An intensive and thorough investigation for all known sorts of failures took place during the shutdown in 2009 and resulted in the largest fraction of enabled HDIs. The tireless efforts during the past shutdowns allowed re-enabling of HDIs and led to an all-time high number of enabled HDIs.\\%
The test-stand was also used for detailed firmware studies in order to improve signal-to-noise (S/N) for the sensors controlled and readout by the SVX-IIe type of chips. The pedestal distribution for the 'old' firmware and the 'new' (improved) firmware is shown in Figure \ref{fig:pedestals_new}a) and b).
By moving certain activities on control lines to a different point in time a significant reduction of the noise level was achieved. The biggest impact in terms of reducing the noise was achieved by moving control signals for 'PreAmp'-reset and 'RampReference'-select further away from the start of digitization. For n-side type of sensors the noise was reduced by approximately 20\% whereas for the p-side type sensors noise level was stable. The noise source is not coupling in the same way to all channels as it can be seen in Figure \ref{fig:pedestals_new}a). Our interpretation is that the control signal pulse generates noise on the chip. The previously persistent second band structure is now completely removed as it can be seen by comparing Figure \ref{fig:pedestals_new}a) and b). This firmware is now used for the entire SMT.\\%
The \dzero data acquisition (DAQ) is a buffered system and consequently the dead-time or front-end busy rate (FEB) %
\begin{figure}[ht]
    \centerline{\includegraphics[width=0.7\columnwidth]{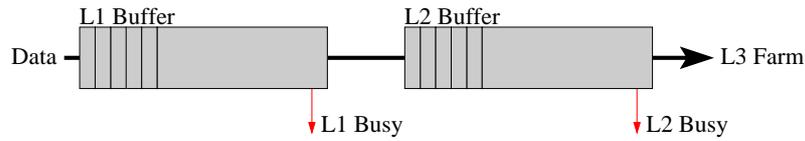}}
  \caption{\label{fig:d0_daq} Simplified sketch of the data flow from left to right. The red arrows indicate a busy signal at different levels if no free buffer is available.}
 \end{figure}
is driven by the amount of data and the ability to process it. Figure \ref{fig:d0_daq} shows a simplified sketch of the data flow in the \dzero experiment. Data are processed by means of a multi-level trigger system (L1, L2, L3). The red arrows indicate a busy signal at different levels if no free buffer is available. \\
Individual SMT crates showed a very peculiar FEB pattern: one would expect that the SMT crate leading in FEB is given by highest data processing load as it takes more time to process more data. Instead the FEB leading SMT crate seems to appear randomly as shown in Figure \ref{fig:feb_pattern}a). It shows FEB rates [\%] of all SMT crates (different colors) as a function of time with the two crates showing the peculiar FEB pattern 
highlighted by the red circles. This happened on an apparently random basis but more frequently at higher trigger rates. The buffer 
handling is organized by a VME read-out board controller (VRBC) \cite{vrbcBoard} which controls the VME read-out boards 
(VRB) \cite{vrbcBoard}, which in turn are gathering the data from the sequencer level as sketched in Figure \ref{fig:hardwareChain}. 
The VRBC firmware was extended with monitoring capabilities for buffer management.
The two plots in Figure \ref{fig:feb_pattern}b) show
 the available buffers (blue), buffers waiting for L2 (green) or L3 decisions (red) as a function of time. The buffer distribution is 
shown for the two SMT crates which exhibit the increased FEB rate (SMT crates 0x65 and 0x67) as highlighted by the red ellipses and arrows.
 A good correlation between the number of available buffers and the FEB was seen. In general there are less available buffers at higher 
trigger rates. The red circles connected by arrows in Figure \ref{fig:feb_pattern}a)-b) highlight the peculiar FEB pattern shown by two 
different SMT crates. This effect was due to the sudden reduction of available buffers (blue) causing increased dead-times for the affected 
SMT crates. Figure \ref{fig:feb_pattern}c) shows the FEB rates [\%] of various detector subsystem crates (SMT crates are colored in red). In addition the global \dzero L1 busy rate consisting of all L1 subsystems is plotted (green). The yellow ellipses highlight an increase of the global L1 busy rate caused by raised FEB rates of particular SMT crates. This illustrates how the sudden reduction of available buffers in SMT crates affected the global L1 busy rate. At higher trigger rates (around $-50$ minutes) the effect is not large. There is an increase of the global L1 busy rate by approximately 2\% at the same time as the jump in FEB for a SMT crate: from 10\% to approximately 12\%. At lower trigger rates (around $-5$ minutes) the effect is smaller and the global L1 busy rate increases only by about 0.4\%. The latter can be understood as the reduced number of buffers has largest impact at high data taking rates.
\begin{wrapfigure}{r}{0.70\textwidth}
  \begin{center}
    \includegraphics[width=0.695\textwidth]{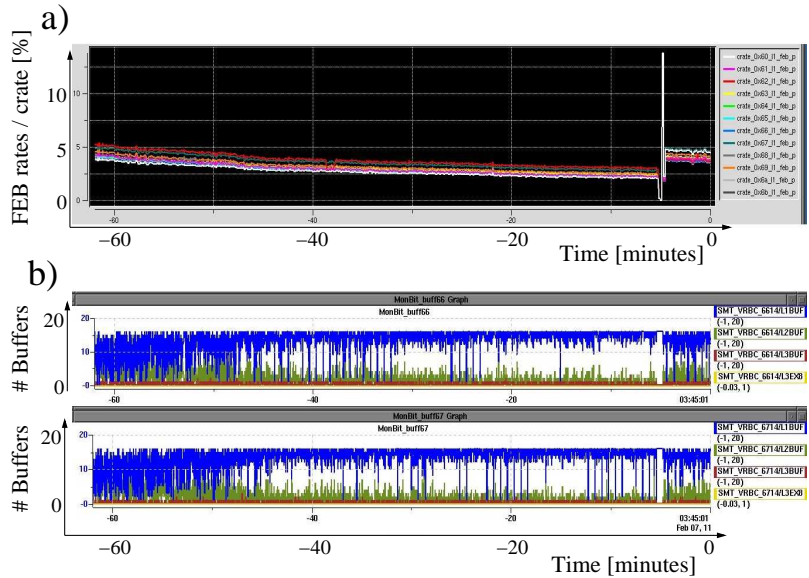}
  \end{center}
   \caption{\label{fig:feb_solved} a) shows the FEB rates [\%] of all SMT crates (different colors) as a function of time with the improved buffer handling firmware. The two plots in b) show again available buffers (blue), buffers waiting for L2 (green) or L3 decisions (red) as an example for two different SMT crates. More details in the text.}
\end{wrapfigure}%
The SMT test-stand allowed tests at high rates of new versions of the VRBC firmware handling buffer management. A more robust 
VRBC firmware version was developed and it did not show this problem anymore. Monitoring data for this modified VRBC firmware version are shown in Figure \ref{fig:feb_solved}a)-b). a) shows the FEB rates [\%] of all SMT crates (different colors) as a function of time with the improved buffer handling firmware. There are no SMT crates showing a significantly higher FEB rate. The increased FEB rates visible at the end of the distribution was due to a change in prescale settings, which increased the event rates. Figure \ref{fig:feb_solved}b) shows the available buffers (blue), buffers waiting for L2 (green) or L3 decisions (red) as a function of time for two different SMT crates. There are no sudden drops in the number of available buffers anymore.

\section{Conclusions}
The SMT has been operated since 2001. Its performance and efficiency have been enhanced using new tools such as the '\rdzero' 
units. The SMT test-stand is a unique piece of equipment to train new experts as well as to reproduce and 
understand subtle effects in the SMT while minimizing impact on global data taking. Three examples have 
been presented: HDI recovery effort, optimization of signal-to-noise and the buffer management problem. In each case the results from the 
test-stand led to improved performance for the entire SMT system. The training of new experts at the test-stand allowed for new 
insights into the operation of the SMT, which in turn increased the stability and performance of the SMT. \\
The SMT detector is performing very well, providing good tracking and vertexing capabilities for the \dzero experiment, which is vital for high efficiency b-tagging and electron/photon identification.

\end{document}